\documentclass[aps,pra,preprint,superscriptaddress,longbibliography]{revtex4-1}
\usepackage{amsmath,amsthm,amsfonts,amssymb,xcolor,graphicx,tikz,color,longtable,array,dsfont}
\usepackage[pdftex]{hyperref} %para ligar las referencias

\begin{document}
	
\title{Lasing in para-Fermi class-B microring resonator arrays}

\author{A. Padr\'on-God\'inez}
\affiliation{Instituto Nacional de Astrof\'{\i}sica, \'Optica y Electr\'onica, Calle Luis Enrique Erro No. 1, Sta. Ma. Tonantzintla, Pue. CP 72840, Mexico.}
\affiliation{Instituto de Ciencias Aplicadas y Tecnolog\'ia, Universidad Nacional Aut\'onoma de M\'exico, Circuito Exterior S/N, Coyoac\'an, AP 04510, Cd. Universitaria, Mexico.}
	
\author{B. Jaramillo-\'Avila}
\affiliation{CONACYT-Instituto Nacional de Astrof\'{i}sica, \'{O}ptica y Electr\'{o}nica, Calle Luis Enrique Erro No. 1. Sta. Ma. Tonantzintla, Pue. CP 72840, Mexico.}
	
\author{B. M. Rodr\'iguez-Lara}
\affiliation{Tecnologico de Monterrey, Escuela de Ingenier\'ia y Ciencias, Ave. Eugenio Garza Sada 2501, Monterrey, N.L., Mexico, 64849.}
\affiliation{Instituto Nacional de Astrof\'{\i}sica, \'Optica y Electr\'onica, Calle Luis Enrique Erro No. 1, Sta. Ma. Tonantzintla, Pue. CP 72840, Mexico.}

\date{\today}

\begin{abstract}
	We demonstrate lasing in arrays of microring resonators with underlying para-Fermi symmetry. 
	The properties of the algebra allow the analytic prediction of propagation constants and normal modes of an array of passive resonators that show the optical analogue of a zero-energy mode that is its own chiral pair.
	The rest of the normal modes in the linear model form chiral pairs that are evenly distributed around the pseudo zero-energy mode. 
	We use this information to construct a class-B laser model where even and odd sites are driven with different strength in a pattern following para-Fermi algebra chirality to demonstrate lasing with enhancement or suppression of the zero-energy mode.
	The former leads to lasing in the zero-energy mode for a large set of driving ratios and is mostly independent of initial field configurations. 
	In contrast, the latter can lead to vanishing fields or lasing with either steady or strongly fluctuating fields depending on the driving ratios and initial field configurations.
\end{abstract}

\maketitle
%--------------------------------------------------------------------------------
%%%%%%%%%%%%%%%%%%%%%%%%%% body %%%%%%%%%%%%%%%%%%%%%%%%%%

%%%%%%%%%%%%%%%%%%%%%%%%%%
\section{Introduction}
%%%%%%%%%%%%%%%%%%%%%%%%%%

Microfabrication techniques enable the production of coupled optical elements where symmetries or topological properties produce novel or desirable behavior. 
This includes the robust generation of particular protected modes and selection mechanisms. 
The introduction of active elements further enhances the possibilities of these devices, where discrete symmetries allow for a plethora of effects; for example, 
mode protection in systems with charge conjugation symmetry \cite{Malzard2015,Ge2017}, 
unidirectionality in systems with parity-time reversal symmetry \cite{Ramezani2010,Feng2013,Yin2013}, 
edge states \cite{Hafezi2013,Poli2015,Parto2018,Zhao2018,Ozdemir2019} and 
flat bands \cite{Leykam2013,Leykam2018,Leykam2018b,Longhi2018,Longhi2019} due to topological effects. 

Coupled mode theory provides a tractable framework to study time evolution or spatial propagation in arrays of coupled elements \cite{Snyder1972,McIntyre1973,Huang1994,Little1997,Liu2005}.
For example, an array of microring resonators with the same curvature secures a symmetric, positive definite mode coupling matrix \cite{Liu2005}.
Then, a change in the reference frame, equivalent to adding a common phase to the fields in the array, can displace the coupled mode matrix diagonal by an arbitrary real constant.
In this sense, pseudo zero-energy modes, that is, modes with vanishing effective propagation constant, are only so in a particular reference frame. 
However, when a discrete symmetry is added to the system, new constraints are added to the coupled mode matrix and, in consequence, to the spectrum of effective propagation constants \cite{Miri2013,Longhi2015,NodalStevens2018,Jaramillo2019,Lumer2019,Zhong2019}. 
Enforcing Hermitian parity \cite{Malzard2015,Ge2017} produces a real spectrum composed of pairs of opposite-sign propagation constants; a pseudo zero-energy mode may appear, for example, in non-degenerate, odd-dimensional systems.
A non-Hermitian parity symmetry \cite{Smirnova2019} produces a complex spectrum with pairs of opposite-sign propagation constants; in other words, the system shows effective source or sink behavior thanks to each chiral pair.
In contrast, charge conjugation \cite{Malzard2015,Ge2017} produces pairs of effective modes with opposite signs in the real part of their propagation constants but the same sign in their imaginary parts; this means identical source or sink behavior in each conjugate pair. 
A system with parity-time symmetry \cite{ElGanainy2007,Makris2008,Makris2008b,Feng2013,Yin2013,HuertaMorales2016,NodalStevens2018} has three scenarios: 
conserved symmetry with real spectrum, 
devil point with dimension-one collapsed spectrum and 
broken symmetry with imaginary spectrum of complex-conjugate pairs. 

Here, we exploit the symmetry effects of para-particle oscillators \cite{HuertaAlderete2017a,HuertaAlderete2017b,HuertaAlderete2018,RodriguezWalton2019}.
We focus on arrays of active, nonlinear microring resonators in an configuration mimicking para-Fermi oscillators described by odd-dimensional, bi-symmetric coupled mode matrices. 
This approach benefits from the rich dynamics offered by coupled rate models \cite{Wang1988,Winful1992,Yanchuk2004}.
We briefly discuss Plyushchay representation of the para-Fermi algebra, Section \ref{sec:Plyuschay}, and the optical para-Fermi oscillator, Section \ref{sec:Oscillator}, to show that, in the passive, linear limit, its spectrum yields a pseudo zero-energy mode that is its own chiral pair, together with chiral pairs of normal modes, due to its odd-dimension and bi-symmetry. 
This informs our proposal of a driving scheme with different pump rates \cite{Johnson2013,Kominis2017,Adams2017,Kominis2020} over the nonlinear array following a pattern prescribed by the parity operator of the algebra, Section \ref{sec:Laser}, to demonstrate either enhancement or suppression of lasing in the zero-energy mode, Section \ref{sec:Numerics}. 
In our case symmetric properties, instead of topology, produce mode selectivity and, hopefully, furthers our symmetry based optical design program \cite{RodriguezLara2015b,RodriguezLara2018}.
We close with our conclusions. 

%%%%%%%%%%%%%%%%%%%%%%%%%%
\section{Para-Fermi algebra} \label{sec:Plyuschay}
%%%%%%%%%%%%%%%%%%%%%%%%%%

A para-Fermi algebra of finite even order admits a representation \cite{Plyushchay1997},
\begin{align}
	[ \hat{I}_{+}, \hat{I}_{-} ] = 2 \hat{I}_{0}\, \hat{\Pi}, \qquad
	[ \hat{I}_{0}, \hat{I}_{\pm} ] = \pm \hat{I}_{\pm},
\end{align}
reminiscent of the angular momentum algebra $su(2)$ with the addition of a parity operator $\hat{\Pi}$ \cite{Macfarlane1993,Dunne1995}.
In this representation, the action of the algebra elements in terms of the $2p+1$ eigenstates of $\hat{I}_{0}$,
\begin{align}
	\hat{I}_{0} |p,n\rangle = n |p,n\rangle,
\end{align}
with $n = -p, -p+1, \ldots, p-1, p$, are 
\begin{align}
	\hat{I}_{+} |p,n\rangle &= \phi(p,n+1)|p,n+1\rangle, \\
	\hat{I}_{-} |p,n\rangle &= \phi(p,n) |p,n-1\rangle, \\
	\hat{\Pi} |p,n\rangle &= (-1)^{p+n} |p,n\rangle.
\end{align}
The structure function,
\begin{align}
	\phi(p,n) = \sqrt{ \left(p+\frac{1}{2}\right) + \left( n - \frac{1}{2} \right)(-1)^{p+n} },
\end{align}
vanishes when the raising (lowering) operator $\hat{I}_{+}$ ($\hat{I}_{-}$) reaches the upper (lower) extreme of the representation, $\phi(p,p+1) = 0$ [$\phi(p,-p) = 0$]. 

Mapping this finite basis into the standard orthonormal basis for a complex vector space of dimension $2p+1$, we can introduce a matrix representation for the para-Fermi algebra operators, where the number and parity operators have diagonal form, 
\begin{align}
	\left[\mathbf{I}_{0}\right]_{i,j} &= (p+1-i) ~\delta_{i,j}, \\
	\left[\mathbf{\Pi} \right]_{i,j} &= (-1)^{i+1} ~\delta_{i,j},
\end{align}
where $i,j=1,\ldots,2p+1$ and the raising and lowering operator have upper and lower diagonal form,
\begin{align}
	\left[\mathbf{I}_{+}\right]_{i,j} &= \phi(p, p+1-i) ~\delta_{i+1,j}, \\
	\left[\mathbf{I}_{-}\right]_{i,j} &= \phi(p, p+1-j) ~\delta_{i+1,j},
\end{align}
in that order. 

%%%%%%%%%%%%%%%%%%%%%%%%%%
\section{Optical para-Fermi oscillator} \label{sec:Oscillator}
%%%%%%%%%%%%%%%%%%%%%%%%%%

\begin{figure}
	\includegraphics[scale=1.5]{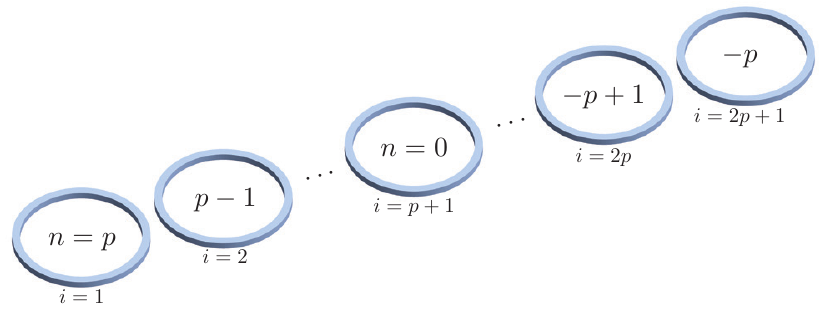}
	\caption{Microring resonator array for the classical simulation of a para-Fermi oscillator.}\label{fig:1}
\end{figure}

We can follow the idea for the optical simulation of para-Fermi oscillators in arrays of coupled waveguides \cite{RodriguezWalton2019} to study lasing in an equivalent array of coupled microring resonators, Fig.~\ref{fig:1}. 
Assuming identical microrings, the coupled mode matrix describing a linear array under first neighbor couplings is given in the following,
\begin{align}\label{eq:cmm}
	\mathbf{M} = \frac{\kappa}{2}~\left( \mathbf{I}_{+} + \mathbf{I}_{-} \right),
\end{align}
where $\kappa$ is a reference coupling constant controlled by the separation between nearest neighbors. 
The eigenvectors and eigenvalues of this matrix are well known \cite{HuertaAlderete2018,RodriguezWalton2019}.
There is an effective pseudo zero-energy mode, 
\begin{align}
	\mathbf{M} \cdot \vec{\mathbf{m}}_{0} = 0 ~\vec{\mathbf{m}}_{0}, 
\end{align}
and the rest of the normal modes, 
\begin{align}
	\mathbf{M} \cdot \vec{\mathbf{m}}_{\pm j} = m_{\pm j} ~\vec{\mathbf{m}}_{\pm j}, \qquad
	m_{\pm j} = \pm \kappa \sqrt{j}, 
\end{align}
with $j=1,\ldots,p$, form chiral pairs, 
\begin{align}
	\mathbf{\Pi} \cdot \vec{\mathbf{m}}_{+j} = \vec{\mathbf{m}}_{-j},
\end{align}
that are evenly distributed around the zero-energy mode which is its own chiral pair,
\begin{align}
	\mathbf{\Pi} \cdot \vec{\mathbf{m}}_{0} = \vec{\mathbf{m}}_{0}.
\end{align}
Thus, a linear gain/loss model that follows the parity operator will keep the zero-energy mode as normal mode,
\begin{align}
	\left(\mathbf{M} + i \gamma	\mathbf{\Pi} \right) \cdot \vec{\mathbf{m}}_{0} = i \gamma \vec{\mathbf{m}}_{0},
\end{align}
while mixing each chiral pair, 
\begin{align}
	\left(\mathbf{M} + i \gamma	\mathbf{\Pi} \right) \cdot \vec{\mathbf{q}}_{\pm j} = q_{\pm j} ~\vec{\mathbf{q}}_{\pm j},
\end{align}
leading to new pairs of eigenvalues and eigenvectors \cite{RodriguezWalton2019},
\begin{align}
	q_{\pm j} &= \pm \sqrt{m_{j}^{2}-\gamma^{2}},
	\nonumber \\
	\vec{\mathbf{q}}_{\pm j} &= N_{\pm j} \left[ \left( m_{j} + q_{\pm j} \right) \vec{\mathbf{m}}_{j} + i \gamma \vec{\mathbf{m}}_{-j} \right],
\end{align}
where $j=1,\ldots,p$ and $N_{\pm j}$ is a normalization constant. 
The eigenmodes of the linear non-Hermitian system, $\vec{\mathbf{m}}_{0}$ and $\vec{\mathbf{q}}_{j}$, have a hierarchy. 
The zero-energy mode has the strongest non-Hermitian behavior, 
with gains (losses) when $\gamma$ is positive (negative), 
as the magnitude of the zero-energy mode imaginary part is the largest of all the modes. 
The rest of the modes have ordered non-Hermitian behavior, 
that of $q_{\pm 1}$ is stronger than that of $q_{\pm 2}$, and so on because the magnitude of their imaginary parts fulfills
$ | \Im (q_{\pm 1}) | \geq | \Im (q_{\pm 2}) | \geq \cdots \geq | \Im (q_{\pm p}) | \ge 0$. 
These pairs of modes suffer an ordered sequence of critical points. 
As the magnitude of the Hermiticity breaking parameter $|\gamma|$ increases, there is a critical point on the local manifold after which $q_{\pm 1}$ become imaginary. 
Subsequently, for a larger value of $|\gamma|$, $q_{\pm 2}$ undergoes the same process, and so on \cite{RodriguezWalton2019}. 
The fact that the zero-energy mode has dominant non-Hermitian behavior means that this mode can either be enhanced or suppressed with respect to the rest of the modes. 
We explore this together with a nonlinear model of coupled class-B microring resonator lasers. 

%%%%%%%%%%%%%%%%%%%%%%%%%%
\section{Class-B laser model} \label{sec:Laser}
%%%%%%%%%%%%%%%%%%%%%%%%%%

In order to take advantage of the linear non-Hermitian behavior, we extend our model to a coupled array of class-B laser microrings, Fig.~\ref{fig:1}.
Here, the state of the $i$-th ring is described by its electric field amplitude $\mathcal{E}_{i}(t)$ and carrier density normalized to transparency $n_{i}(t)$ with $i=1,\ldots,2p+1$ .
The dynamics of these are governed by the following nonlinear differential equation set,
\begin{flalign}\label{eq:pflaser}
	i\, \dot{\mathcal{E}}_{i}(t) 
	=& 
	\frac{i(1-i\alpha)}{2} \left\{ -\frac{1}{\tau_{p}} + \sigma \left[n_{i}(t)-1\right] \right\} \mathcal{E}_{i}(t) 
	\nonumber \\
	&+ \sum_{j=1}^{2p+1} \mathbf{M}_{i,j}\, \mathcal{E}_{j}(t), 
	\\
	\dot{n}_{i}(t)
	=&
	R_{i} - \frac{n_{i}(t)}{\tau_{s}} - \frac{2\left[n_{i}(t)-1\right]}{\tau_{s}} \left| \mathcal{E}_{i}(t) \right|^{2},
\end{flalign}
where the linewidth enhancement factor $\alpha$, carrier and cavity lifetimes $\tau_{s}$ and $\tau_{p}$ and differential gain proportionality $\sigma$ are constant parameters characterizing the identical rings. 
An isolated ring pumped above its threshold rate $R_{(th)} = [1+1/(\sigma\,\tau_{p})]/\tau_{s}$ displays lasing. 
At long times, $t\gg \tau_{s}, \tau_{p}$, the electric field at the isolated ring reaches a finite value which may be steady or fluctuating depending on the ring parameters and initial conditions. 
In contrast, if the pump rate is below threshold, the field evolves to zero. 

We require driving that follows the parity operator pattern; that is, even (odd) sites share the same pump rate, 
\begin{align}
	R_{m} 
	=
	R_{(th)} 
	\left\{
	\begin{array}{ll}
		1 + \Delta_{E} & \quad \text{for even } i, \\
		1 - \Delta_{O} & \quad \text{for odd } i.
	\end{array}
	\right.
\end{align}
where the real, dimensionless parameters $\Delta_{E}, \Delta_{O}$ fix even and odd pump rates. 
This class-B laser model, with the para-Fermi oscillator at its core, should provide a similar behavior to the linear non-Hermitian case where the zero-energy mode is enhanced (suppressed) with the addition of effective relative gain at even (odd) sites \cite{RodriguezWalton2019}.
In the following, we numerically explore the two alternatives offered by our nonlinear model. 
For zero-energy mode enhancement, we use pump rates above (below) lasing threshold at even (odd) microrings; that is, positive dimensionless parameters $\Delta_{E}, \Delta_{O} > 0$ are the nonlinear analogue to having $\gamma > 0$ in the linear gain/loss model \cite{Ge2016}. 
We seek zero-energy mode suppression using the opposite configuration; pump rates above (below) lasing threshold at odd (even) microrings providing $\Delta_{E}, \Delta_{O} < 0$ equivalent to $\gamma < 0$ in the linear non-Hermitian model. 

%%%%%%%%%%%%%%%%%%%%%%%%%%
\section{Numerical analysis} \label{sec:Numerics}
%%%%%%%%%%%%%%%%%%%%%%%%%%

\begin{figure*}
	\centering
	\includegraphics{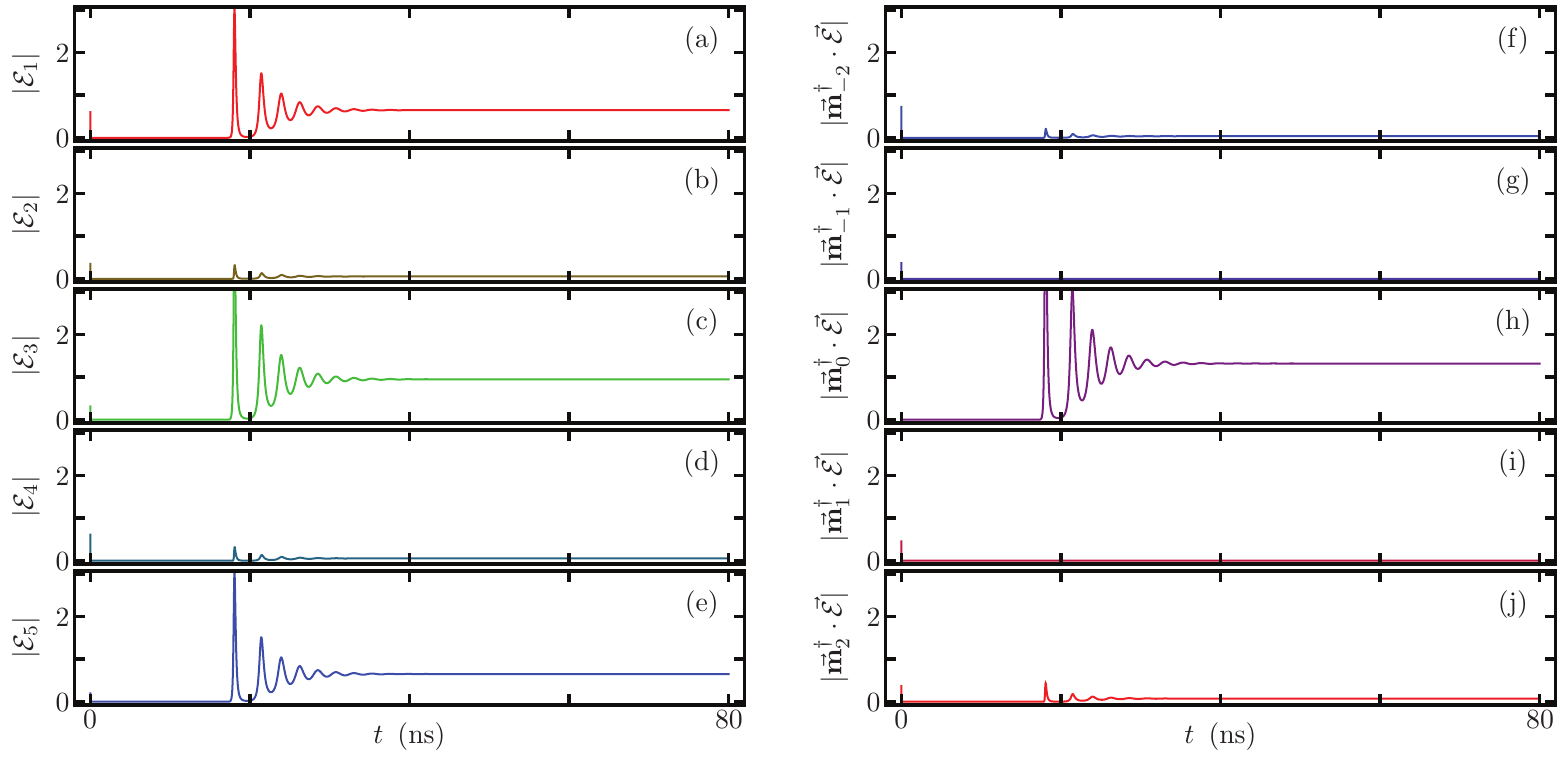}
	\caption{Time evolution of the absolute value of (a)--(e) localized field amplitudes in a five ring array in the zero-energy mode enhancement scheme and (f)--(j) their projection into the normal modes provided by the linear para-Fermi oscillator with (h) being the projection into the zero-energy mode.}\label{fig:2}
\end{figure*}

Our numerical simulation follows the experimental realization of topological active arrays \cite{Parto2018}. 
That is, an array of identical microring resonators with linewidth enhancement factor $\alpha = 3$, carrier and cavity lifetimes $\tau_{s} = 4~\mathrm{ns}$ and $\tau_{p} = \tau_{s}/100$, and differential gain proportionality constant $\sigma = 24/\tau_{p}$.
In order to realize the para-Fermi oscillator, we set a reference coupling constant to $\kappa=10^{11}~\mathrm{Hz}$ corresponding to a separation of roughly $177~\mathrm{nm}$ between the borders of the rings.
We report two configurations. 
One with five microrings with separations 
$d_{1,2}=d_{4,5}=177~\mathrm{nm}$; 
$d_{2,3}=d_{3,4}=216~\mathrm{nm}$. 
Another with eleven microrings, 
$d_{1,2}=d_{10,11}=142~\mathrm{nm}$; 
$d_{2,3}=d_{ 9,10}=216~\mathrm{nm}$; 
$d_{3,4}=d_{ 8, 9}=149~\mathrm{nm}$; 
$d_{4,5}=d_{ 7, 8}=177~\mathrm{nm}$; 
$d_{5,6}=d_{ 6, 7}=160~\mathrm{nm}$. 
All these separations are within reported experimental values for arrays of coupled microring resonators \cite{Parto2018,Zhao2018}.
We report the average of a thousand realizations considering random normalized initial conditions for the analysis of mode enhancement/suppression. 
We explore zero-energy mode enhancement/suppression and whether field magnitudes are steady or fluctuating for an average of 50 random normalized initial conditions for a matrix of 1600 points in the phase space defined by the pump rates in even and odd sites $\left( \Delta_{E}, \Delta_{O} \right)$.

In the zero-energy mode enhancement scheme, even (odd) rings are driven $5\%$ above (below) threshold; that is, $\Delta_{E} = \Delta_{O} = 0.05$.
We set the initial carrier densities to zero $n_{i}(0) = 0$ and generate random normalized initial complex fields amplitudes without projection into the zero-energy mode of the linear para-Fermi oscillator.
Figure \ref{fig:2} displays the time evolution of just one numerical experiment. 
The left column, Fig.~\ref{fig:2}(a) to Fig.~\ref{fig:2}(e), shows the time evolution of the absolute value of the field amplitude at each of the microrings, $\vert \mathcal{E}_{i}(t) \vert$.
The right column, Fig.~\ref{fig:2}(f) to Fig.~\ref{fig:2}(j), shows the projection onto the modes of the linear array, $\vert \vec{\mathbf{m}}_{j} \cdot \vec{\mathbf{\mathcal{E}}}(t) \vert$.
As expected under this setup and initial conditions, the array lases in a stable mode with a field distribution equivalent to that of the zero-energy mode of the linear model, Fig.~\ref{fig:2}(h).

\begin{figure}
	\includegraphics{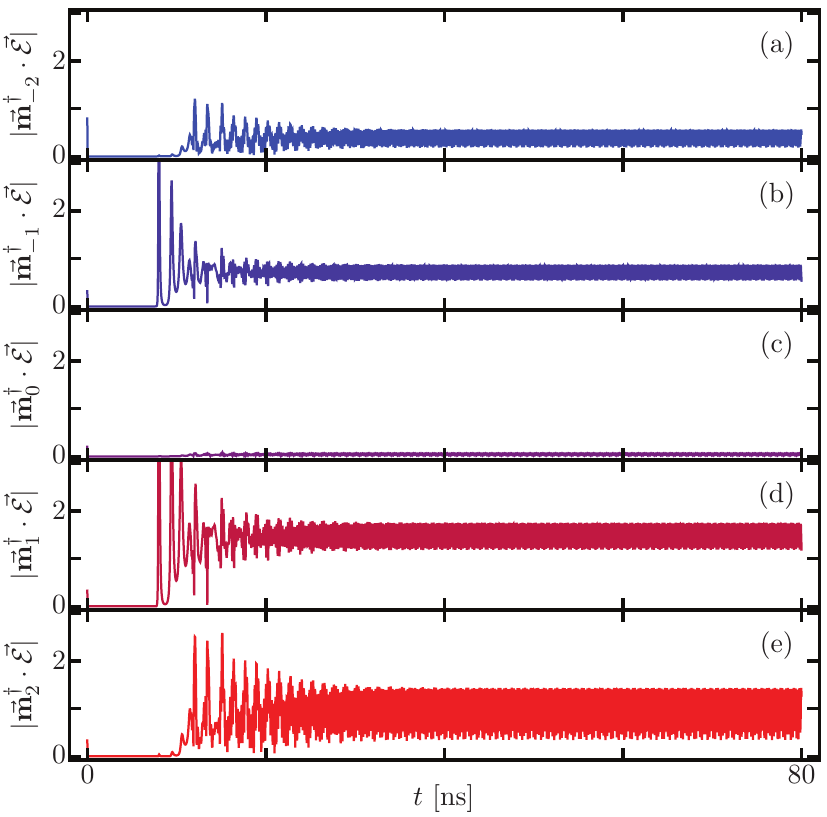}
	\caption{Time evolution of the absolute value of the field amplitudes projection into the normal modes provided by the linear para-Fermi oscillator in a five ring array with the zero-energy mode suppression scheme.}\label{fig:3}
\end{figure}

Figure \ref{fig:3} shows results for the zero-energy mode suppression scheme, where odd (even) rings are driven $30\%$ above (below) the threshold rate; that is, 
$\Delta_{E} = \Delta_{O} = -0.30$.
Again, we set the initial carrier densities to zero $n_{i}(0) = 0$.
The initial fields have now completely random normalized complex amplitudes.
Under this setup and initial conditions, the field and their projections onto the para-Fermi oscillator normal modes show strong fluctuations. These may arise due to a stable limit cycle or to unstable behaviors.
The zero-energy mode is clearly suppressed with respect to the rest of the normal modes, Fig.~\ref{fig:3}(c).

\begin{figure}
	\includegraphics[scale=1.5]{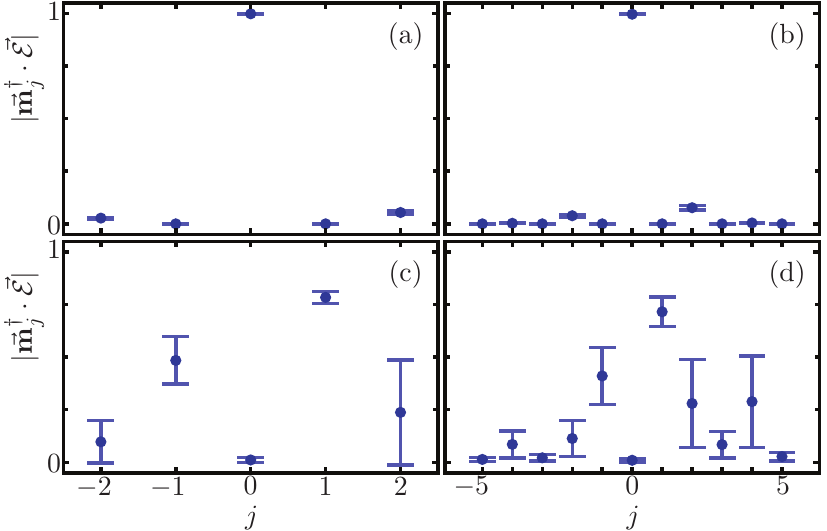}
	\caption{Statistical analysis for zero-energy mode (a)--(b) enhancement and (c)--(d) suppression schemes. The dots show the average absolute value of the long-time fields onto the normal modes of the para-Fermi oscillator and the bars show the standard deviation of a thousand random initial samples once they are renormalized. The left column shows results for an array with five microrings and the right column for eleven microrings.}\label{fig:4}
\end{figure}

In order to conduct a statistical analysis of these schemes, we calculate the evolution of a thousand different random initial field configurations with the parameters and conditions mentioned above.
For each random numerical experiment, we calculate the central value of the projection onto each normal mode of the para-Fermi oscillator at long times $ t > 15 \tau_{s}$ and renormalize the field evolution using this value.
We use these central values to calculate the mean and standard deviation of a thousand experiments, Fig.~\ref{fig:4}.
The scheme for zero-energy mode enhancement shows steady fields with a large zero-energy mode component, Fig.~\ref{fig:4}(a) and Fig.~\ref{fig:4}(b).
The suppression scheme, provides lasing configurations with negligible zero-energy mode components that are strongly fluctuating, Fig.~\ref{fig:4}(c) and Fig.~\ref{fig:4}(d); that is, the field amplitudes at each microring present fast oscillations as shown in Fig.~\ref{fig:3}.
In a manner similar to the hierarchical ordering of modes in the linear non-Hermitian model, lasing in the modes $j = \pm 1$ tend to dominate. 
By running over a large sample of different normalized initial field configurations with $\Delta_{E} = \Delta_{O} = 0.05$, we find that initial field configurations of that magnitude lead to stationary fields in the zero-energy mode, see Fig. \ref{fig:4}(a)--(b). 
Similarly, with $\Delta_{E} = \Delta_{O} = -0.30$, different normalized initial conditions lead to fluctuating fields with zero projection to the zero-energy mode, see Fig. \ref{fig:4}(c)--(d).

\begin{figure}
	\includegraphics[scale=1.5]{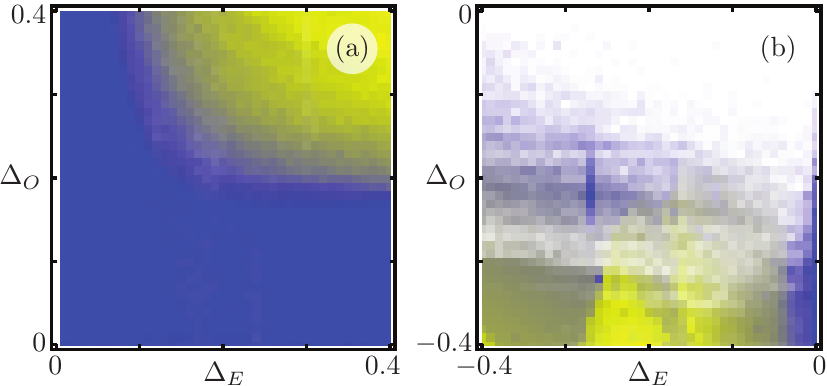}
	\caption{(Color online) Pump rate map defined by even and odd site rates, $\left(\Delta_{E},\Delta_{O}\right)$ in that order. 
		Each point is the averaged result of fifty random, normalized initial field configurations for zero-energy mode (a) enhancement and (b) suppression schemes. 
		Blue (dark) corresponds to steady fields and yellow (light) to fluctuating fields. 
		Fading to white is proportional to the number of non-lasing vanishing fields occurrences.}\label{fig:5}
\end{figure}

We can explore the pump rate effect at even and odd microrings using pump rates, $\Delta_{E} \neq \Delta_{O}$. 
As long as both $\Delta_{E}$ and $\Delta_{O}$ are positive (negative), we remain in the zero-energy mode enhancement (suppression) scheme. 
Figure \ref{fig:5} displays results for (a) enhancement and (b) suppression schemes using the microring parameters mentioned before. 
For each pair of pump rate values $\left(\Delta_{E},\Delta_{O}\right)$, we use a set of fifty random normalized complex initial field configurations and calculate their long time behavior $t > 15 \tau_{s}$. 
From these results, we use the standard deviation of the long-time fields at all the microrings of each sample as measurement of whether the field magnitudes are steady or fluctuating and we use the ratio of initial field configurations that produce non-lasing vanishing fields as a measure for lasing.
The zero-energy mode enhancement scheme yields steady fields for small values of $\left(\Delta_{E},\Delta_{O}\right)$, blue (dark) regions in Fig.~\ref{fig:5}(a).
As the rates grow, the field magnitudes become strongly fluctuating, yellow (light) regions in Fig.~\ref{fig:5}(a).
The suppression scheme, shows vanishing fields for small values of $\left(\Delta_{E},\Delta_{O}\right)$, white region in Fig.~\ref{fig:5}(b), and mostly fluctuating field magnitudes, yellow (light) fading to white region, with a small steady field area, blue (dark) fading to white region in Fig.~\ref{fig:5}(b). 
This suggests a complicated phase structure for zero-energy mode suppression schemes where paths to chaos may be explored.
As the driving changes from zero-energy mode enhancement, Fig. \ref{fig:5}(a), to suppression, Fig. \ref{fig:5}(b), the dynamics displays a discontinuity in the sense that, in the former case, the system evolves towards finite, steady fields. In the latter, the system evolves towards vanishing fields.

%%%%%%%%%%%%%%%%%%%%%%%%%%%%%%
\section{Conclusion}
%%%%%%%%%%%%%%%%%%%%%%%%%%%%%%

We propose arrays of coupled active microrings, each one a class-B laser, that realize para-Fermi oscillators. 
This optical simulation of the para-Fermi algebra allows for a pseudo zero-energy mode that is its own chiral pair; that is, an eigenstate of the parity operator.
We use this fact to set individual microring pump rates according to their position in the array, following a pattern provided by the parity operator of the para-Fermi algebra. 
This driving scheme allows for two scenarios: enhancement or suppression of the zero-energy mode. 

Zero-energy mode enhancement shows little sensitivity to the initial field configuration. 
At long evolution times $t \gg \tau_{s}, \tau_{p}$, even (odd) site pump rates slightly above (below) threshold produce steady fields. Increasing (decreasing) these pump rates away from threshold leads to fluctuating fields.
In contrast, the zero-energy mode suppression scheme is highly sensitive to the initial fields in the microrings. 
At long evolution times, even (odd) site pump rates slightly below (above) threshold produce vanishing fields. 
As pump rates deviate from threshold, most of the initial configurations produce strongly fluctuating fields. 

We find important to note that these phenomena, zero-energy mode enhancement or suppression, arise as a result of the algebraic properties and symmetries of the system, that translate into finite odd-dimensional microring arrays, instead of topological effects. 
Symmetries are a powerful tool to generate and control protected states or modes in arrays with a small number of active optical elements. 

%%%%%%%%%%%%%%%%%%%%%%% References %%%%%%%%%%%%%%%%%%%%%%%%%
%%%%\section*{References}
%\bibliography{ref-PM-Rings}

\end{document}